# LANDSCAPE PHAGES AND THEIR STRIPPED PROTEINS AS SELF ASSEMBLING AFFINITY REAGENTS FOR MEDICINE AND TECHNOLOGY

*Valery A. Petrenko[1], Jennifer R. Brigati[1], Jennifer Sykora[1], Eric V. Olsen[1], Iryna B, Sorokulova[1], Galina A. Kouzmitcheva[1], I-Hsuan Chen[1], James M. Barbaree[2], Bryan A. Chin[3] and Vitaly J. Vodyanoy[1]*

[1]College of Veterinary Medicine, [2]Department of Biological Sciences, and [3]Department of Mechanical Engineering, Auburn University, Auburn, AL, U.S.A.

## ABSTRACT

Filamentous phage, such as fd used in this study, are thread-shaped bacterial viruses. Their outer coat is a tube formed by thousands equal copies of the major coat protein pVIII. We constructed libraries of random peptides fused to all pVIII domains and selected phages that act as probes specific for a panel of test antigens and biological threat agents. Because the viral carrier is infective, phage borne bio-selective probes can be cloned individually and propagated indefinitely without needs of their chemical synthesis or reconstructing. We demonstrated the feasibility of using landscape phages and their stripped fusion proteins as new bioselective materials that combine unique characteristics of affinity reagents and self assembling membrane proteins. Biorecognition layers fabricated from phage-derived probes bind biological agents and generate detectable signals. The performance of phage-derived materials as biorecognition films was illustrated by detection of streptavidin-coated beads, *Bacillus anthracis* spores and *Salmonella typhimurium* cells. With further refinement, the phage-derived analytical platforms for detecting and monitoring of numerous threat agents may be developed, since the biodetector films may be obtained from landscape phages selected against any bacteria, virus or toxin. As elements of field-use detectors, they are superior to antibodies, since they are inexpensive, highly specific and strong binders, resistant to high temperatures and environmental stresses.

## 1. INTRODUCTION. PHAGE AS BIOSELECTIVE NANOMATERIAL

The filamentous bacteriophages Ff (fd, f1 and M13) are long, thin viruses, which consist of a single-stranded circular DNA packed in a cylindrical shell composed of the major coat protein pVIII (98%), and a few copies of the minor coat proteins capping the ends of the phage particle (Fig. 1). The protein pVIII is a typical membrane protein. During infection of the host *Escherichia coli*, the phage coat is dissolved in the bacterial cytoplasmic membrane, while viral DNA enters the cytoplasm. The protein is synthesized in infected cell as a water-soluble cytoplasmic precursor, which contains an additional leader sequence of 23 residues at its N-terminus. When this protein is inserted into the membrane, the leader sequence is cleaved off by a leader peptidase. Later,

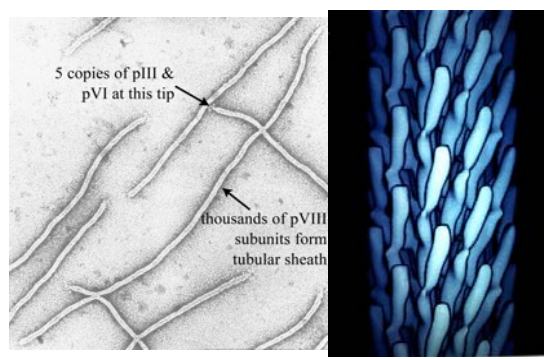

Fig.1. Filamentous phage. Left: electron micrograph. The aminoterminus of pIII proteins are visible and pointed by arrow. Right: Segment of ~1% of phage virion with the array of pVIII proteins shown as electron densities (Micrograph and model cortesy of Irina Davidovich, Gregory Kishchenko and and Lee Makowski)

during the phage assembly, the newly synthesized pVIII proteins are transferred from the membrane into the coat of the emerging phage. Thus, the major coat protein can change its conformation to accommodate to various distinctly different forms of the phage and its precursors: phage filament, intermediate particle (I-form), spheroid (S-form), and membrane-bound form discussed in this report. This structural flexibility of the major coat protein pVIII is determined by its unique architecture, which is studied in much detail.

Foreign peptides were displayed on the pVIII protein soon after pIII display was pioneered [1,2] The pVIII-fusion phages display the guest peptide on every pVIII subunit, increasing the virion's total mass by up to 20%. Yet, remarkably, such particles can retain their ability to infect *E. coli* and form phage progeny. Such particles were eventually given the name "landscape phage" to emphasize the dramatic change in surface

*Valery A. Petrenko, Jennifer R. Brigati, Jennifer Sykora, Eric V. Olsen, Iryna B, Sorokulova, Galina A. Kouzmitcheva, I-Hsuan Chen, James M. Barbaree, Bryan A. Chin and Vitaly J. Vodyanoy*

LANDSCAPE PHAGES AND THEIR STRIPPED PROTEINS AS SELF ASSEMBLING AFFINITY REAGENTS

architecture caused by arraying thousands of copies of the guest peptide in a dense, repeating pattern around the tubular capsid, as illustrated by Fig. 2 [3]. It was shown that the foreign peptides replacing three or four mobile amino acids close to the N-terminus of the wild-type protein pVIII don't disturb considerably the general architecture of virions and don't change the conformation of the fused pVIII protein in membranes. However, the foreign peptides decorate the phage body creating defined organic surface structures (landscapes) that varies from one phage clone to the next. A landscape library is a huge population of such phages, encompassing billions of clones with different surface structures and biophysical properties [3]. Therefore, the landscape phage is unique micro-fibrous material that can be selected in the affinity binding protocol. The binding peptide comprising up to 20% of the phage mass and up to 50% of the phage surface may be easily prepared by cultivation of the infected bacteria and isolation of the secreted phage particles by precipitation. Landscape phages have been shown to serve as substitutes for antibodies against various antigens and receptors [4-7] including live bacterial cells [8,9,13], gene-delivery vehicles [10] and detection probes in biosensors [11,12,14].

Fig. 2. Structure of landscape phage. Foreign peptides are pictured with dark atoms; their overall arrangement corresponds to the model of Marvin, 1994.

Although landscape phages themselves are very attractive bioselective filamentous materials, their alternative forms, such as biorecognition films may be more advantageous in some technical and medical applications where bioselective materials are required. As discussed above, ability of the major coat protein pVIII to form bioselective films emerges from its intrinsic function as membrane protein. We explored these remarkable properties of the pVIII protein for preparation of the filamentous phage probes and bioselective films of the fusion phage proteins, which are discussed in this report.

## 2. LANDSCAPE PHAGE DISPLAY LIBRARIES

The first large ($10^9$-clone) landscape library was constructed by splicing the underlined degenerate coding sequence below into the beginning of the pVIII coat-protein gene, replacing wild-type codons 2–4 [3]

```
DNA    GCAGNKNNKNNKNNKNNKNNKNNGGATCCCGCAAAAGC
pVIII   A  X  X  X  X  X  X  X  X  D  P  A  K  A
        1                          5  6  7  8  9
```

where each N represents an equal mixture of all four nucleotides (A, G, C and T) and K—a mixture of G and T. As a result of this modification, every pVIII subunit in a phage is five amino acids longer than wild-type, and displays a "random" sequence of eight amino acids: $X_8$ above. In any single clone, the random octamer is the same in every particle, but almost every clone displays a different random octamer. The octamers are arranged regularly around the outside of the virion, occupying a substantial fraction of the surface, as shown on the Fig. 2. In another, 9-mer library, aspartic acid D in Position 5 of the major coat protein (see above) is also deleted and substituted with random nonamers. Another, f8-alpha library [5], presented by $10^8$ clones, have random amino acids in the amphipathic helix (positions 12-13, 15-17 and 19) of the major coat protein pVIII, as shown below:

```
1 2 3 4 5 6 7 8 9 10 11 12 13 14 15 16 17 18 19 20 …
A E G E D P A K A A  F  X  X  L  X  X  X  A  X  E …
```

where X stands for any amino acid.

To increase diversity of potential probes, we plan to design a new set of phage libraries with different modes of peptide display. More specifically, we propose a new type of phage landscape libraries with modified amphiphilic domain of the pVIII protein, in which different mutations of amino acids 12-19 increase conformational diversity of foreign peptides inserted at the N-terminus of pVIII protein. These highly diverse hypothetical libraries will serve as additional rich source of landscape phage probes.

## 3. THE LANDSCAPE PHAGES AS SUBSTITUTE ANTIBODIES

Landscape phages were explored as substitute antibodies using a set of model antigens: streptavidin from *Streptomyces avidinii*, avidin from chicken egg white and β-galactosidase from *Escherichia coli* [4]. Each protein antigen was absorbed to the surface of a 35-mm polystyrene Petri dish and the dish reacted with the landscape library. Unbound phage virions were washed away, and bound phage eluted with acid buffer and amplified by infecting fresh bacterial host cells. After three rounds of selection, individual phage clones were propagated and sequenced partly to determine the amino acid sequence of the displayed peptide. Binding of the selected phage to their respective antigens was characterized by enzyme linked immuno sorbent assay (ELISA) and by quartz crystal microbalance (QCM) in which the phages immobilized on the plastic surface of the ELISA wells or gold electrodes of QCM reacted with their antigens in solution phase. These tests demonstrated specific dose-dependent binding of each antigen to the phage it has selected. Inhibition ELISAs and QCM

*Valery A. Petrenko, Jennifer R. Brigati, Jennifer Sykora, Eric V. Olsen, Iryna B, Sorokulova, Galina A. Kouzmitcheva, I-Hsuan Chen, James M. Barbaree, Bryan A. Chin and Vitaly J. Vodyanoy*

LANDSCAPE PHAGES AND THEIR STRIPPED PROTEINS AS SELF ASSEMBLING AFFINITY REAGENTS

measurements verified that non-immobilized peptide-bearing phage compete with immobilized phage for binding to their respective antigens [4,5,12]. These experiments with different antigens have shown that landscape phages may be used as a new type of substitute antibodies—filaments that can bind protein and glycoprotein antigens with nanomolar affinities and high specificity.

Using the landscape phage libraries, which were characterized in the model experiments, and advanced selection protocols we isolated phages that bind strongly and specifically to live bacterial and tumor cells [6-13], as exemplified in subsection 6.

## 4. STRIPPED FUSION PHAGE PROTEINS AS BIORECOGNITION PROBES

In the model experiments, the filamentous landscape phages displaying streptavidin- and *Salmonella typhimurium*-binding peptides fused to all 4,000 copies of the major coat protein pVIII were converted to a new biorecognition affinity reagent—"stripped phage" [3,4]. The stripped phage is a composition of disassembled phage coat proteins with dominated (98%) recombinant major coat protein pVIII that forms bioselective vesicles with unique landscape of target-binding peptides. In our examples, the stripped phages were prepared by treatment of the landscape phages with chloroform followed by transformation of resulted spheroids into the biorecognition films by their reconstruction with phospholipids. In experiments with streptavidin binders it was demonstrated by competition ELISA that the stripped phage proteins retain the target-binding properties of the selected phage. Specific binding of the spheroid to the target cells was visualized by transmition electron microscopy (Fig. 3). The stripped phage proteins were also characterized by there binding to the target cells using quartz crystal balance (QCM) technique and scanning electron microscopy (Subsection 6).

## 5. ROBUSTNESS OF THE PHAGE PROBES

Most detection devices have traditionally relied on antibodies as diagnostic probes. Their use outside of a laboratory, however, may be problematic because antibodies are often unstable in severe environmental conditions. Environmental monitoring requires thermostable probes, such as landscape phage carrying thousands of foreign peptides on their surface, that are superior to antibodies and can operate in non-controlled conditions. While parent wild-type phages are known to be extremely stable in various media at high temperatures, no work has been done to demonstrate stability of landscape phage probes. We examined the thermostability of a landscape phage probe and a monoclonal antibody specific for β-galactosidase in parallel in an ELISA format [14]. They were both stable for greater than 6 months at room temperature, but at higher temperatures the antibody degraded more rapidly than the phage probe. As shown in the Fig.4 phage retained detectable binding ability for more than 6 weeks at 63°C, and 3 days at 76°C. The activation energy of phage degradation was determined to be 31,987 cal/mol. These results confirm that phage probes are highly thermostable and can function even after exposure to high temperatures during shipping, storage and operation.

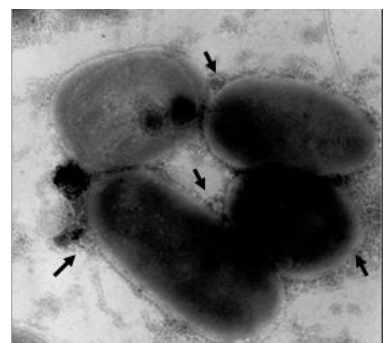

Fig. 3. Binding of spheroids to the target cells. Arrows point at some areas of massive accumulation of spheroids at the cell surface.

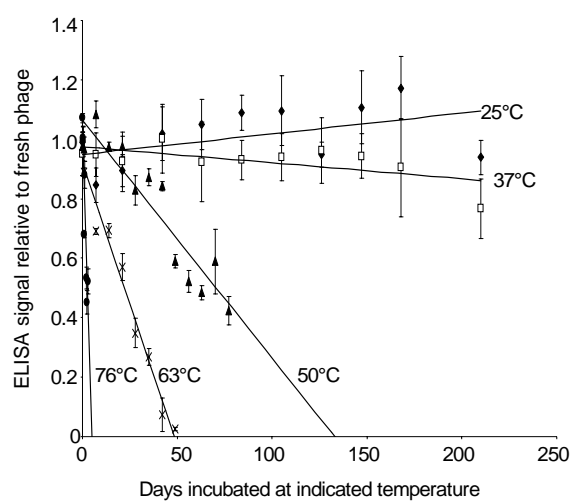

Fig. 4. Thermal stability of the phage. Phage fixed to the wells of EIA plates was incubated at 25°C (♦), 37°C (□), 50°C (▲), 63°C (X), and 76°C (●). At various intervals, plates were removed from the incubator, fresh phage was added to an unused set of wells, and β-galactosidase was allowed to bind to the phage. β-galactosidase binding was visualized by the addition of substrate ONPG. ELISA signal values (mOD/min) are relative to the signals in the wells containing fresh phage. Error bars are indicative of standard deviations of triplicate wells

*Valery A. Petrenko, Jennifer R. Brigati, Jennifer Sykora, Eric V. Olsen, Iryna B, Sorokulova, Galina A. Kouzmitcheva, I-Hsuan Chen, James M. Barbaree, Bryan A. Chin and Vitaly J. Vodyanoy*

## 6. LANDSCAPE PHAGES AS DETECTION PROBES

We demonstrated that the phage landscape libraries, in which random foreign peptides are fused to all pVIII domains (Fig. 2), contain many potential probes for surface markers of cells, spores and bacteria. Phage probes against biological threat agents, such as *Bacillus anthracis* spores and *Salmonella typhimurium* were isolated in a nonbiased multistage selection procedure using immobilized spores or bacteria as a selector. The performance of the probes in detection of these threats was illustrated by a precipitation test, enzyme-linked immunosorbent assay (ELISA), fluorescence-activated cell sorting, magneto-strictive sensors, and fluorescent, optical and electron microscopy. Representative landscape phage selected with *B. anthracis* spores bind to the selector strain at a higher level than to other species of *Bacillus* spores (Fig. 6). Similarly, the phage selected with *S. typhimurium* binds more strongly to the selector in comparison with nine other gram-negative bacteria, predominately *Enterobacteriaceae* (Fig. 7). A small amount of cross reactivity of this phage was noted with *Yersinia enterocolitica and Citrobacter freundii.* The complex of phage with bacteria was visualized by fluorescence microscopy (not shown) and transmission electron microscopy (TEM) (Fig. 5), demonstrating the multivalent character of phage-bacteria binding**.**

To serve as detection probes phage were immobilization onto the sensor surfaces by one of the following ways:

- Phage self-assemblage on Langmuir-Blodgett (LB) phospholipid by biotin/streptavidin coupling;
- Direct physical adsorption of phage to the sensor surface; or
- "Phage stripping" — coating of the sensor with phage-derived fusion peptide probes.

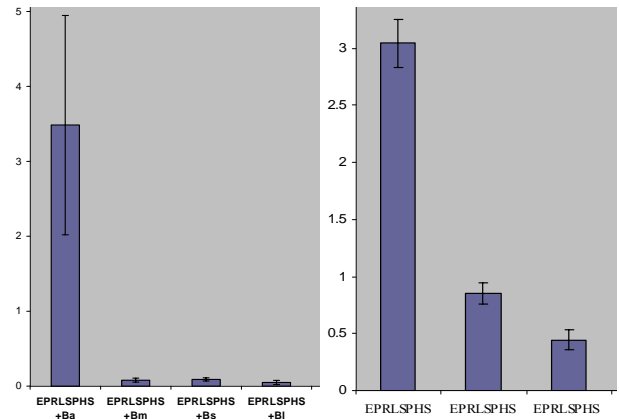

Fig. 6. Binding of selected phage EPRLSPHS to *Bacillus* spores in comparison with *B. megaterium*, *B. subtilis*, *B. licheniformis,. B. cereus*, and *B.thuringiensis* in the coprecipitation assay. x-axis: species of spores that were mixed with selected phage; y-axis: percent of phage recovered by coprecipitation with spores

In the LB method monolayers containing biotinylated phospholipids were transferred onto the gold surface of acoustic wave sensors and treated with streptavidin and biotinylated phage. The phage-loaded sensor demonstrated specific dose-dependent binding of β-galactosidase from *E.coli* [9]. It was observed that the affinity of the complex depends on the mode of phage immobilization and type of analytical platform: 0.6 nM by acoustic wave sensor versus 30 nM by ELISA. The difference in affinities was attributed to the monovalent (ELISA) and divalent (sensor) interaction of the phage with β-galactosidase, as was indicated by the analysis of

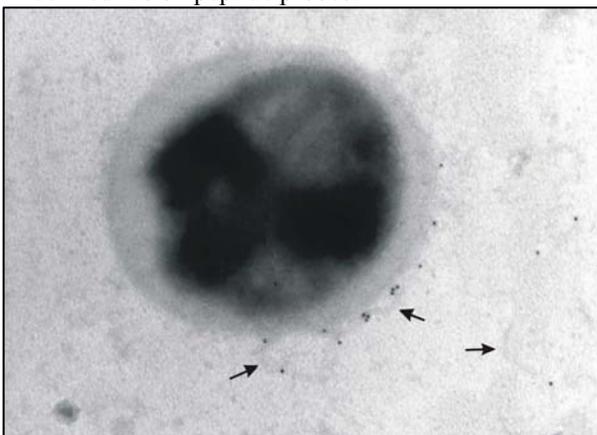

Fig. 5: TEM micrograph of bacteria-phage complex. Phage is labeled with gold nanoparticles (arrows). Adapted from (Petrenko and Sorokulova, 2004)

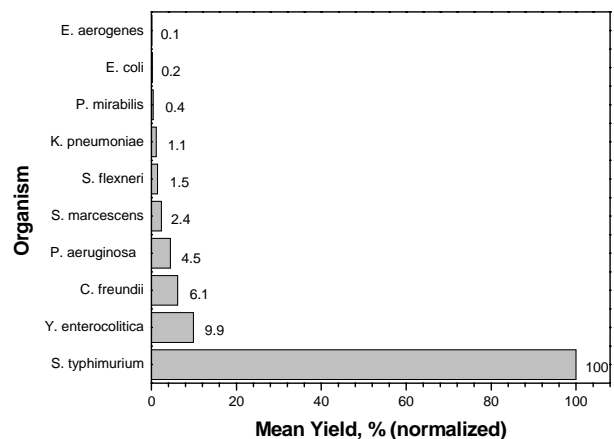

Fig. 7. Selectivity of phage VTPPTQHQ as determined by coprecipitation assay. Mean yield (output/input x 100) percentage is the average of three separate experiments normalized to the maximal mean yield of 3.4% from *S. typhimurium* (1). Numbers above bars indicate respective percentages.

*Valery A. Petrenko, Jennifer R. Brigati, Jennifer Sykora, Eric V. Olsen, Iryna B, Sorokulova, Galina A. Kouzmitcheva, I-Hsuan Chen, James M. Barbaree, Bryan A. Chin and Vitaly J. Vodyanoy*

LANDSCAPE PHAGES AND THEIR STRIPPED PROTEINS AS SELF ASSEMBLING AFFINITY REAGENTS

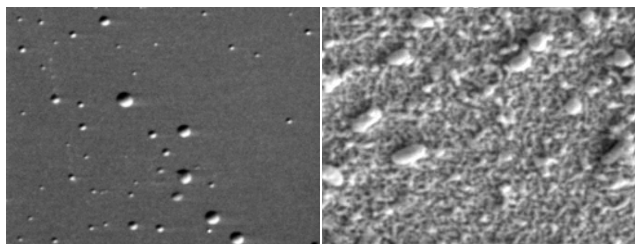

Fig. 7. Study of interaction of *S. typhimurium* ATCC 13311 with selected phage VTPPTQHQ by scanning electron microscopy: Left – plain gold surface nontreated with phage and bacteria; Right -- gold surface, coated with phage and treated with bacteria (the arrow shows an attached *Salmonella* cell)

binding curves using the Hill presentation. One or another mode of interaction probably depends on the conformational freedom of the phage immobilized to the solid surface. Binding of the phage is quite specific because the response is reduced by 85% if β-galactosidase is preincubated with 4 nM phage. Binding of the phage to β-galactosidase is very selective: presence of 1000-fold excess of bovine serum albumin in mixture with β-galactosidase does not considerably change the ELISA signal and reduces the biosensor signal only by 4%.

We found that phage can adsorb directly onto the gold surfaces [11]. In these experiments, the acoustic wave sensor (Maxtek) with gold electrodes was exposed to phage in suspension. Following an incubation period, the sensor was rinsed in water and tested with analytes. A sensor for β-galactosidase showed the value of $EC_{50}$ of approximately 2 nM, what is comparable with results obtained by the above described self-assembling LB method. Another biosensors specific for *S. typhimurium* demonstrated a linear dose-response relationship over six decades of bacterial concentration. Scanning electron microscopy (SEM) (Fig. 7) confirmed bacterial binding to the sensor. The sensitivity of the biosensor (-10.9 Hz) was vastly greater than the established background. The lower limit of detection based on the dose-response curve was estimated at 100 cells/ml.

Bioselective films of the stripped fusion phage proteins were prepared by conversion of the phage filaments into spheroids [3,4], their rupture, compression of the formed layers and their transfer onto the sensor surface using the LB technique [15]. We have shown that fusion phage proteins produce a functional biospecific coating. For example, Fig. 8 demonstrates signals generated by acoustic wave sensors coated with the film formed from phage specific for streptavidin. For each streptavidin-coated bead (~1 μm *d*.) concentration ($10^4$-$10^8$ particles/ml) reacted with sensors, the signal approaches a steady-state response within 500 s. The interaction of the beads with the phage fusion peptides is specific since the signal is significantly lower for beads coated with BSA (lower line - circles). Binding of the beads to the sensor was confirmed by SEM.

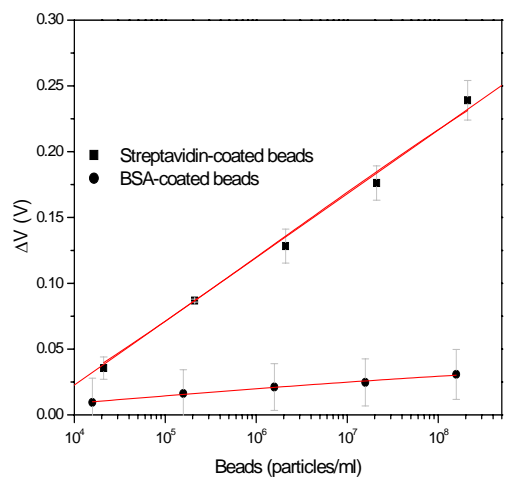

Fig. 8. Differential dose-response of steady-state output sensor voltages as a function of streptavidin– (■) and BSA–coated (●) bead concentrations.

The stripped phage proteins were used for design of a biodetector for *S. typhimurium*. It was prepared from spheroids combined with phospholipid–1,2-diphytanoyl-sn-glycero-3-phospho-choline. The majority of the sensors demonstrated satisfactory mean sensitivity, greater than 2.5 mV/decade. These results demonstrate proof in concept development of biosensors that incorporate phage as probes for the detection of threat agents such as *S. typhimurium.*

## 7. CONCLUSION

The presented data show that the phage engineering, which grounds on the natural mechanisms of selection, amplification and self assemblage, is a powerful and very precise technique that allows directed nanofabrication of bioselective materials, with possible applications to biosorbents, biosensors, nanoelectronics, and other areas of medicine, technology, and environmental monitoring. In particular, the genetically driven "phage landscaping" allows the generation of libraries possessing diverse nanostructures accommodated on the phage's surface – a huge resource of diagnostic and detection probes. Biorecognition layers fabricated from the selected phage-derived probes bind biological agents, and as a part of analytical platforms, generate detectable signals. They may be suitable as robust and inexpensive antibody substitutes for field-use detectors and real time monitoring devices for critical threat agents.

*Valery A. Petrenko, Jennifer R. Brigati, Jennifer Sykora, Eric V. Olsen, Iryna B, Sorokulova, Galina A. Kouzmitcheva, I-Hsuan Chen, James M. Barbaree, Bryan A. Chin and Vitaly J. Vodyanoy*

LANDSCAPE PHAGES AND THEIR STRIPPED PROTEINS AS SELF ASSEMBLING AFFINITY REAGENTS

## 8. REFERENCES

[1] Ilyichev, A. A., Minenkova, O. O., Tatkov, S. I., Karpyshev, N. N., Eroshkin, A. M., Petrenko, V. A., and Sandakhchiev, L. S. (1989). Construction of M13 viable bacteriophage with the insert of foreign peptides into the major coat protein. Doklady Biochemistry (ProcAcad Sci Ussr)-EnglTr *307*, 196-198.

[2] Petrenko, V. A., and Smith, G. P. (2005). Vectors and Modes of Display. In Phage Display in Biotechnology and Drug Discovery, S. S. Sidhu, ed. (Bo Raton, FL, U.S.A., CRC Pressw, Taylor & Francis Group), pp. 714 pp. 63-110.

[3] Petrenko, V.A., Smith, G.P., Gong, X., Quinn, T., 1996. A library of organic landscapes on filamentous phage. Protein Eng. 9, 797-801.

[4] Petrenko V.A., Smith, G.P., 2000. Phage from landscape libraries as substitute antibodies. Protein Eng. 13, 589-592.

[5] Petrenko, V.A., Smith, G.P., Mazooji, M.M., Quinn, T., 2002. Alpha-helically constrained phage display library. Protein Eng. 15, 943-950.

[6] Romanov, V. I., Durand, D. B., and Petrenko, V. A. (2001). Phage display selection of peptides that affect prostate carcinoma cells attachment and invasion. Prostate *47*, 239-251.

[7] Samoylova, T. I., Petrenko, V. A., Morrison, N. E., Globa, L. P., Baker, H. J., and Cox, N. R. (2003). Phage probes for malignant glial cells. Molecular Cancer Therapeutic *2*, 1129-1137.

[8] Brigati J., Williams, D.D., Sorokulova, I.B., Nanduri, V., Chen, I-H., Turnbough, C.L., Petrenko, V.A., 2004. Diagnostic probes for *Bacillus anthracis* spores selected from a landscape phage library. Clin. Chem. 50, 1899-1906.

[9] Petrenko V.A., Sorokulova, I.B., 2004. Detection of biological threat agents. A challenge for combinatorial biochemistry. J. Microbiol. Methods 58, 147-168.

[10] Mount, J. D., Samoylova, T. I., Morrison, N. E., Cox, N. R., Baker, H. J., and Petrenko, V. A. (2004). Cell Targeted Phagemid Rescued by Pre-Selected Landscape Phage. Gene *341,* 59-65.

[11] Olsen E.V., Sorokulova I.B., Petrenko V.A., Chen I-H., Barbaree J.M., and V.J. Vodyanoy (2005). Affinity-selected filamentous bacteriophage as a probe for acoustic wave biodetectors of *Salmonella typhimurium.* Biosensors and Bioelectronics. Aug 4; [Epub ahead of print]

[12] Petrenko V.A., Vodyanoy, V.J., 2003. Phage display for detection of biological threat agents. J. Microbiol. Methods 53, 253-262.

[13] Sorokulova, I.B., Olsen, E.V., Chen, I-H., Fiebor, B., Barbaree, J.M., Vodyanoy, V.J., Chin, B.A., Petrenko, V.A., 2005. Landscape Phage Probes for *Salmonella typhimurium.* J. Microbiol. Methods (In Press).

[14] Brigati J.R. and V.A. Petrenko, 2005. Thermo-stability of Landscape Phage Probes. Analytical and Bioanalytical Chemistry V.382, p.1346-1350.

[15] Petrenko, V.A., Vodyanoy, V.J., Sykora, J.C. Methods of forming monolayers of phage-derived products and uses thereof. U.S. Patent Application. Methods of Forming Monolaers of Phage-Derived Products and Uses Thereof. Appl. No. 10/792,187; Filed March 3, 2004.

*ACKNOWLEDGMENT*

This work was supported by ARO/DARPA grant #DAAD 19-01-10454 (VAP), NIH NIH-1 grant #R21 AI055645 (VAP) and USDA grant #99-34394-7546 (Dr. Bryan Chin). We are grateful to Dr. M.A. Toivio-Kinnucan (College of Veterinary Medicine, Auburn University) for excellent electron microscopic support.